\documentclass[10pt,conference]{IEEEtran}

\usepackage{graphicx}
\usepackage{amsfonts}
\usepackage{dsfont}

\usepackage{amsmath}

\usepackage{algorithm}
\usepackage{algpseudocode}

\usepackage{xcolor}
\newcommand{\yuan}[1]{}
\newcommand{\comment}[1]{}
\usepackage[normalem]{ulem}
\newcommand{\fixes}[1]{}

\usepackage{multirow}
\usepackage{hyperref}

\usepackage{caption}
\begin{document}

\title{Chain-of-Experts (CoE): Reverse Engineering Software Bills of Materials for JavaScript Application Bundles through Code Clone Search}

% \author{Anonymous submission}

\author{
\IEEEauthorblockN{Leo Song\IEEEauthorrefmark{1}, Steven H. H. Ding\IEEEauthorrefmark{2}, Yuan Tian\IEEEauthorrefmark{1}, Li Tao Li\IEEEauthorrefmark{1}, Philippe Charland\IEEEauthorrefmark{3}, Andrew Walenstein\IEEEauthorrefmark{4}}
%\vspace{0.05in}
\IEEEauthorblockA{\IEEEauthorrefmark{1}Queen's University. \emph{leo.song@queensu.ca}, \emph{y.tian@queensu.ca}, \emph{litao.li@queensu.ca}}
\IEEEauthorblockA{\IEEEauthorrefmark{2}McGill University. \emph{steven.h.ding@mcgill.ca}}
\IEEEauthorblockA{\IEEEauthorrefmark{3}Mission Critical Cyber Security Section, Defence R\&D Canada. \emph{philippe.charland@drdc-rddc.gc.ca}}
\IEEEauthorblockA{\IEEEauthorrefmark{4}Security Research and Development at BlackBerry. \emph{awalenstein@blackberry.com}}
%\vspace{0.05in}
}
% \email{leo.song@queensu.ca}
% \affiliation{
%   \institution{Queen's University}
%   \city{Kingston}
%   \state{Ontario}
%   \country{Canada}
%   \postcode{K7P 0M6}
% }

% \author{Steven H. H. Ding}
% \email{steven.h.ding@mcgill.ca}
% \affiliation{
%   \institution{McGill University}
% }

% \author{Yuan Tian}
% \email{y.tian@queensu.ca}
% \author{Li Tao Li}
% \email{litao.li@queensu.ca}
% \affiliation{
%   \institution{Queen's University}
% }

% \author{Philippe Charland}
% \email{philippe.charland@drdc-rddc.gc.ca}
% \affiliation{
%     \institution{Mission Critical Cyber Security Section, Defence R\&D Canada}
% }

% \author{Andrew Walenstein}
% \email{awalenstein@blackberry.com}
% \affiliation{
%     \institution{Security Research and Development at BlackBerry}
% }
% \renewcommand{\shortauthors}{Song et al.}
% \renewcommand{\shorttitle}{SemSeg}

\maketitle

\begin{abstract}
A Software Bill of Materials (SBoM) is a detailed inventory of all components, libraries, and modules in a software artifact, providing traceability throughout the software supply chain. With the increasing popularity of JavaScript in software engineering due to its dynamic syntax and seamless supply chain integration, the exposure to vulnerabilities and attacks has risen significantly. A JavaScript application bundle, which is a consolidated, symbol-stripped, and optimized assembly of code for deployment purpose. Generating a SBoM from a JavaScript application bundle through a reverse-engineering process ensures the integrity, security, and compliance of the supplier's software release, even without access to the original dependency graphs.

This paper presents the first study on SBoM generation for JavaScript application bundles. We identify three key challenges for this task, i.e., nested code scopes, extremely long sequences, and large retrieval spaces. To address these challenges, we introduce Chain-of-Experts (CoE), a multi-task deep learning model designed to generate SBoMs through three tasks: code segmentation, code classification, and code clone retrieval. We evaluate CoE against individual task-specific solutions on 500 web application bundles with over 66,000 dependencies. Our experimental results demonstrate that CoE offers competitive outcomes with less training and inference time when compared with combined individual task-specific solutions. Consequently, CoE provides the first scalable, efficient, and end-to-end solution for the SBoM generation of real-world JavaScript application bundles.

\end{abstract}


\section{Introduction}
Software Bill of Materials (SBoM) provides a detailed inventory of all components, libraries, and modules used in software, enabling source provenance transparency over the complex software supply chain~\cite{xia2023empirical}.
%{Shift-right security strategies emphasize conducting security assessments on software release artifacts prior to the deployment phase.}
SBoM generation has been an integral part of software supply chain for further security analysis~\cite{harutyunyan2020managing}. 
Recent studies~\cite{bi2023way,stalnaker2024boms,kloeg2024charting, zahan2023software} have highlighted two key challenges in SBoM: the tooling for generating SBoMs for different types of applications and the traceability of dependencies. This traceability involves showing the exact locations of dependencies in the released software, rather than just the metadata catalog provided by the supplier.
% \yuan{not sure "traceability of dependencies" mean what, my understanding is that SBoM is all about this? Meanwhile, how CycloneDX
% and SPDX relate to our study? are they commercial tools and we want to do something similar but open-source? How our format of SBoM differs from them?}

JavaScript applications are prevalent in software supply chain. 
% These applications are bundled from JavaScript source code, which facilities SBoM generation by programming language analysis tools. 
% \yuan{maybe the previous two lines should be moved to a later position after mentioning the popularity of Javascript and high risk in supply chain attack?} 
JavaScript's flexible syntax facilitates rapid development and deployment, enabling developers to swiftly deliver updates to end users.
However, this shift-right agility comes with increased risk, as it may inadvertently introduce security vulnerabilities.
The rising occurrences of supply chain attacks~\cite{supply2021report} reflect similar security concerns in the JavaScript domain~\cite{zahan2022weak, rabbi2024sbom}.

\begin{figure}[t]
    \centering
    \includegraphics[width=\columnwidth]{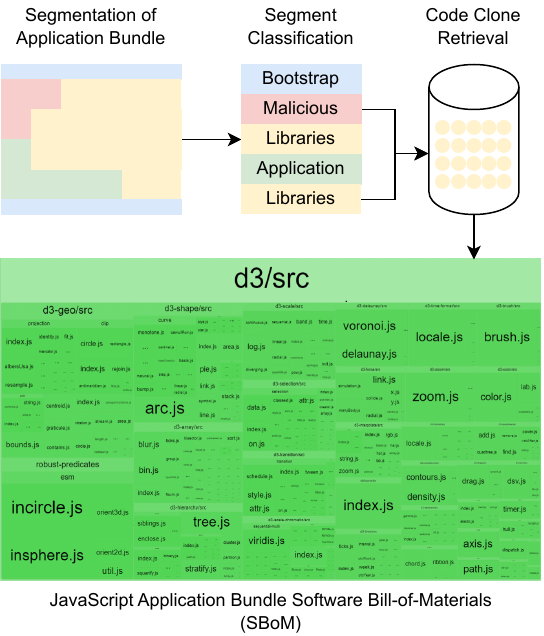}
    \caption{We propose to segment, classify, and the clone search against a JavaScript application bundle file to recover its dependency graph through a reverse engineering process.}
    \label{fig:model_design}
\end{figure}

JavaScript applications are released and deployed as application bundles. An application bundle, distinct from traditional source code and binaries, aggregates the application's script files and includes the source code of all dependencies, into a single, consolidated, symbol-stripped, code-mixed, and optimized assembly file.
Webpack, one of the most popular JavaScript bundlers, sees over 25 million downloads per week at the time of writing this paper. 
Existing JavaScript SBoM generation tools rely on the dependency graph from the source code, such as \textit{package.json}~\cite{rabbi2024sbom}.
These dependency metadata files are typically either unavailable or unreliable through the supply chain, making bundles the only available information for code analysis. 
Our goal is to generate an SBoM from a JavaScript application bundle through a reverse-engineering process (see the input and output of Figure~\ref{fig:model_design}). It ensures the integrity, security, and compliance of the supplier's software release, even without access to the original source code.
% \yuan{are you saying that our solution is not to create new format for SBoM, but just use Javascript bundles as it is, the main contribution is the support of dependency retrieval (in terms of code segment) from it?}
% Webpack, one of the most popular JavaScript bundlers, incurs over 25 million downloads per week at the time of writing this paper.
% As of 2024, over 4.8 million bundles are hosted on the JavaScript package repository (NPM).
% Millions of websites
% Most importantly, the dependency metadata files such as source maps, are typically either unavailable or unreliable during the release stage of software development, making bundles the only available information for code analysis. 

Code clone search has emerged as a promising approach for generating traceable SBoMs from released binary firmware by searching for reused binary code~\cite{kwaners0}. However, directly applying this method to a JavaScript application bundle is not feasible due to the complexity and mixed nature of the code within these bundles.
The bundler optimization process mixes the dependencies and application source code together.
It also removes and replaces all symbols with simpler characters, making it more challenging to analyze.
% and scan the composition of JavaScript applications.
% This complicates the task of dissecting and scanning the composition of JavaScript applications.
Particularly, we identify three challenges in developing the first SBoM solution for JavaScript application bundles:

\textbf{C1: Nested Scopes.} 
Unlike binary files and other programming languages where released artifacts can be easily divided into different units such as functions, JavaScript bundles have complex scopes due to JavaScript's syntax and the optimization process. There doesn't exist established approach to segment the entire JavaScript package into smaller units that can be easily analyzed and searched. Therefore, a segmentation method is needed for JavaScript bundles to determine the minimal code unit of analysis.
% Unlike binary files and other programming languages, JavaScript bundles have complex scopes due to JavaScript's syntax and optimization. As a result, it is difficult to divide the whole JavaScript package into distinct segments to retrieve their sources
% \yuan{shouldn't we mention segmentation is a common step for SBoM generation first?}. 
% In contrast, binary files~\cite{ding2019asm2vec} and other source code files~\cite{svajlenko2014towards} can be easily divided into function blocks for classification, clone detection, and retrieval tasks.

%yuan{why we need to divide JS package into segments?}\comment{covered in next section}

\textbf{C2: Extreme Long Sequence Length.} 
JavaScript packages can contain thousands of identifiers on average, and an application bundle can contain over one million identifiers, mixed with bootstrapping code, bundler-injected code, application source code files, and dependency libraries. This extremely long mixed sequence makes traditional code clone search methods for dependency retrieval not directly applicable, as these methods did not account for code mixing. 

Additionally, certain parts of the bundler, such as bootstrapping code and bundler-injected code, can cause a large number of false positives during code search, which requires classification and filtering. Furthermore, artifact material classification labels, such as libraries and applications, are typically required in standard SBoM formats such as CycloneDX~\cite{cyclonedx}. 
% Without a universal segmentation method, JavaScript packages can contain thousands of tokens on average. Long sequences can lead to poor performance for current retrieval methods. %\yuan{why deep learning methods are needed?}\comment{use the wording retrieval here and talk about methods in next section}
% with limited GPU resources, such as the 512-token restriction for BERT-style pre-trained models.

\textbf{C3: Large Retrieval Space.} As of 2024, the NPM repository contains over 4.8 million files for libraries with different versions. Existing comparison-based code clone detection methods~\cite{zhang2019novel, guo_graphcodebert_2021} require pairwise comparisons between a query and all candidates in the repository, leading to poor efficiency and limited scalability. Instead, embedding-based deep learning methods can leverage Approximate Nearest Neighbor search algorithms, enabling more scalable code clone retrieval without the need for exhaustive pairwise comparisons. 

% Consequently, this vast number of sources makes it impossible to retrieve the clone of a JavaScript segment using current algorithms.

\begin{figure*}[t]
    \centering
    \includegraphics[]{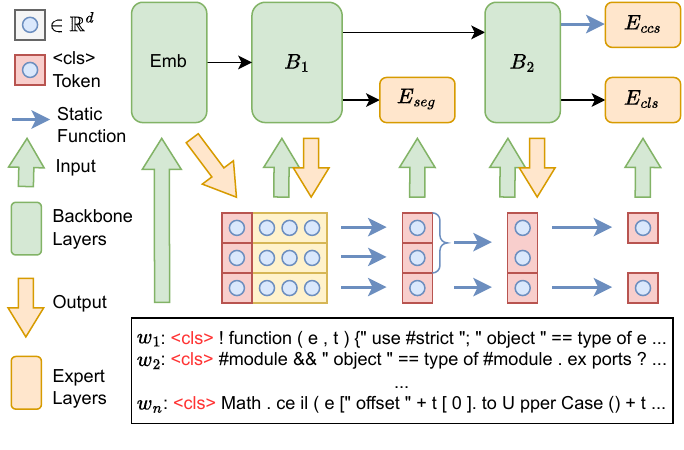}
    \caption{The architectural design and data flow for the CoE model.
    Expert models are positioned according to the order of tasks.
    Each window of the inputs is fed into the first backbone model and the classification expert.
    Class tokens are aggregated based on these predictions and processed through a second backbone model following segmentation.
    This setup facilitates the subsequent tasks of code classification and code clone retrieval, employing a classification expert model and cosine similarity metrics.}
    \label{fig:model}
\end{figure*}

% The SBoM generation process is to list all constituent components, manifesting their sources, versions and any essential metadata. 

Thus, we propose a pipeline with three tasks for JavaScript SBoM, which are code segmentation, code classification, and code clone retrieval through embedding search, as illustrated in Figure~\ref{fig:model_design}.
The code segmentation step divides the complete JavaScript bundle into distinct segments, where each segment represents one source component. Next, the code classification step categorizes each segment, identifying it as third-party library code, application code, or bootstrapping code. 
Classified segment labels help reduce false positives in subsequent clone searches and offer more detailed layout information to the security analyst.
% A type of a component can be a whole application, a third-party library, a framework, and etc.
% This step can identify those segments exhibiting prominent patterns, such as bundler framework bootstrap code.
Then, each segment can be searched to retrieve the sources and their respective versions from the relevant proprietary repositories.
This is conducted at the file level across a repository of JavaScript libraries such as NPM.
This approach facilitates an efficient and accurate search among sources at scale. These tasks collectively contribute to more effective analysis and categorization of JavaScript code for SBoM.
% \yuan{now my understanding is that, each library, framework, whole application can form SBoM using CycloneDX, but they still can not be identified when utilized in another application?}

% SBoM generation for JavaScript application bundles.
% Lack of SBoM tooling is a prominent issue in literature~\cite{hiesgen2022race, supply2021report, zahan2022weak}, and existing JavaScript SBoM generation tools rely on the dependency graph~\cite{rabbi2024sbom}\yuan{duplicate with previous description? in the beginning of this section}.
% We divide the entire SBoM process into three tasks, where deep learning methods have demonstrated remarkable performance.
%\sout{Deep learning methods, particularly those for natural language processing, have shown remarkable performance in understanding and interpreting complex patterns in textual data and programming languages. JavaScript code, with its syntactic and semantic richness, can be effectively parsed and analyzed using these models.}
% The current solutions for natural language segmentation~\cite{koshorek2018text, glavas_two-level_2020, lo2021transformer}, classification~\cite{liu2021topic, guo_graphcodebert_2021}, and embedding~\cite{reimers2019sentence, jiang2023low, behnamghader2024llm2vec}, can be well-applied for these three tasks.
% In addition, some work~\cite{barrow2020joint, inan2022structured} utilizes a Multi-Task learning approach for text segmentation and classification to utilize the semantic contexts of the long sequence inputs, which inspired our methodologies.

Following this three-step pipeline, 
we propose a novel language-model-based deep learning architecture named CoE, as shown in Figure~\ref{fig:model}. The CoE architecture consists of a chain of backbone models and distinct expert models, enabling a continuous flow of information and shared learning across different tasks. The backbone models contain shared parameters that facilitate the downstream learning process across various tasks, while the outputs of each expert model are tailored to specific tasks. The novelty of our approach lies in the complete generation of a SBoM by a single neural network, ensuring efficient and accurate outputs. This is achieved through multi-task learning with joint losses, allowing the model to simultaneously optimize for multiple objectives.
%The architecture of CoE is characterized by a chain of backbone models and distinct expert models, facilitating a seamless flow of information and shared learning across different tasks. The backbone models houses shared parameters for the downstream learning process across various tasks, whereas the outputs of each expert model are tailored to specific tasks.
%The novelty of our algorithms lies in a complete SBoM generation by one neural network, ensuring efficient and accurate outputs. This is achieved by multi-task learning with joint losses.
Our contributions can be summarized as follows:
\begin{enumerate}
\item  We propose the first solution to build the SBoM generation for JavaScript application bundles by dividing it into three distinct tasks: code segmentation, classification, and clone retrieval\footnote{\href{https://github.com/}{GitHub link placeholder}}. The formulation of these tasks is elaborated upon in Section~\ref{sec:formulation}.

\item We propose a novel CoE model to fuse multiple tasks into one end-to-end deep learning model with enhanced performance and less training time. The methodologies are detailed in Section~\ref{sec:methodologies}.
  
\item In our experimental results presented in Section~\ref{sec:experiment}, we evaluate our proposed approaches using real-world JavaScript packages. We also assess the operational time and the deployment feasibility with limited hardware resources.  
\end{enumerate}
Additionally, a thorough review of related methods is conducted in Section~\ref{sec:related_work}, elaborating existing research gaps in this area.
% The source code is available at \url{https://github.com/L1NNA/sem-seg}.

\section{Problem Formulation} \label{sec:formulation}
% \noindent \textbf{Notation.} In the sequel, $x, \textbf{x}, X, \textbf{\textit{X}}, \mathbb{X}$ denote a scalar, vector, array, matrix, and set. We denote $\lVert \cdot \rVert$, $\lVert \cdot \rVert_{ed}$, $|X|$ as the Euclidean Distance, Levenshtein Edit Distance, and array length.
%\yuan{it is better to define the overall goal and then three tasks} \yuan{it sounds like our solution in the end is for clone retrieval. then why is the task SBoM generation?}\comment{fixed}

Given a query \( Q \) (e.g. a file) and some datasets \( \mathbb{D} \)s (e.g. code repositories), the goal of SBoM, by its definition, is to locate all the constituent sub-queries from their proprietaries.
The first task is to partition \( Q \) into sub-queries \( q_i \) using a segmentation function \( seg \).
All queries are non-overlapping and consecutive.
Then, we classify the queries into their destined search pool (i.e. $\mathbb{D}$) by a classification model $cls$.
% Next, we encode the queries and all the elements in $\mathbb{D}$ by the encoding model, named $enc$.
Next, we calculate the similarity between queries and dataset elements using a distance metric $sim$, such as cosine similarity, Euclidean distance, and Hamming distance.
Based on these similarity scores, the code clone retrieval algorithm ranks elements to identify the most similar element (or the top \( k \) nearest neighbors) within \( \mathbb{D} \).
In formal terms:
\begin{align*}
seg(Q) & = (q_1, q_2, \dots, q_s), Q = q_1 || \dots || q_s \\
cls(q) &: q \in \mathbb{D} \\
ccr(q, p) &: sim(q, p) \geq sim(q, p_j), \forall p_j \in \mathbb{D}
\end{align*}
The queries \( Q \), \( q \), and dataset elements \( p \) can reside in the Hamming Space \( \{0, 1\}^n \) when representing binary segments, or in the Euclidean space \( \mathbb{R}^n \) for lexical tokens.

As discussed in the introduction section, we are tackling the SBoM analysis for \textbf{JavaScript} packages.
Consequently, under the scope of this paper, $Q$ is any JavaScript application bundle\yuan{\sout{a file representing the whole JS application bundle?}}, $\mathbb{D}$ is a collection of JavaScript files (e.g. a package management repository), $q, p$ are in $\mathbb{Z}^n$, where $n$ is the number of input tokens.
Since we are leveraging deep learning methodologies, the elements will be further encoded to d-dimensional vectors $\mathbf{q}, \mathbf{p}$ by the neural network \( enc \), such that $\mathbf{q}, \mathbf{p}$ are in \( \mathbb{R}^{d} \).
In addition, we choose the Hierarchical Navigable Small World Graph (HNSW) \cite{8594636} to index $\mathbb{D}$ and cosine similarity as our similarity metric $sim$\yuan{do we need to mention this implementation detail here? could be part of the CoE? in the end, comparing models should ignore which similarity metric used right?}\comment{I'll keep it here and remove the last sentence}.
Thus, our primary focus lies in optimizing three tasks: \textbf{code segmentation}, \textbf{code classification}, and \textbf{code clone retrieval}.
Specifically, we optimize the deep neural network architecture for the segmentation function \( seg \), the classification function \( cls \), and the code clone retrieval algorithm \( ccr \) with cosine similarity loss.

\begin{figure}[htbp]
    \centering
    \includegraphics[width=.8\columnwidth]{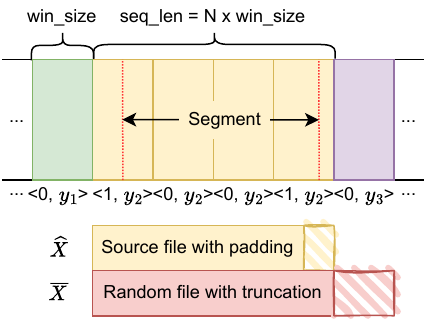}
    \caption{The figure illustrates the inputs and outputs of the sliding window operation. The vertical red dotted lines represent segment boundaries, while different color blocks indicate different segments. The entire input script is divided into $N$ windows, each of size $win\_size$. The number of input tokens for the CoE model, denoted as $seq\_len$, is equal to $N \times win\_size$. Each window is labeled with two pieces of information: the presence of a segment boundary and the class name. Additionally, each window is paired with a source file and a random contrastive file for cosine similarity matching. It is important to note that the pairs must be padded or truncated to match the $seq\_len$ limit.}
    \label{fig:sliding_window}
\end{figure}

\section{Chain-of-Experts Architecture Design} \label{sec:methodologies}
The CoE model essentially tackles three tasks simultaneously.
Figure~\ref{fig:model} illustrates the data flow and its high-level architecture.
The inputs for CoE are code tokens and the outputs are the boundary indicators, classes of each segment, and the vector embedding.
We utilize a multi-level structure with a chain of backbone and expert neural networks.
The backbone models are responsible for semantic encoding, as in $B_1$, and global data decoding, such as $B_2$.
The outputs of each expert neural network are tailored to meet the requirements of each task with matching prediction losses.
In addition, the CoE pipeline includes BPE tokenization and sliding windows for input processing and a novel ``Segmentation Masking''  in $B_2$ layer to emulate the manual segmentation process. The sliding window approach and two-level architecture are expected to tackle the \textbf{Long Sequence Length} challenge. The segmentation expert model and segmentation masking are designed to solve the \textbf{Nested Scopes} challenge.
The joint model and embedding outputs are robust for an embedding task with \textbf{Large Retrieval Space}.

\subsection{Sliding windows} \label{sec:sw}
First, we address the challenge of long JavaScript code sequences by dividing the code into sliding windows, as shown in Figure~\ref{fig:sliding_window}. During the training phase, we label each window with a `1' if it contains a segment boundary and `0' otherwise. Additionally, each window is labeled with its class name for the code classification task and the target file name for the code clone retrieval task. This design enables deep learning models to predict boundary locations within each window while maintaining an acceptable input size. By combining windows that belong to the same segment, we can perform classification and clone retrieval for extended code segments.

\begin{equation} \label{eq:sw}
\begin{aligned}
seq\_len &= N \times win\_size \\
X^i &= [cls] \: || \: Q[i : i + win\_size] \\
X &= X^1 \: || \: ... \: || \: X^N \\
Y^i_{seg} &= \mathds{1}_{\{X_i \text{ has a boundary}\}} \\
Y_{cls} &= y_{cls} \\
Y_{ccr} &= \begin{cases}
    1, \text{ for } cos\_sim(X, \hat{X})\\
    -1, \text{for } cos\_sim(X, \overline{X})
\end{cases} \\
i &= i + step
\end{aligned}
\end{equation}

Equations~\ref{eq:sw} list the inputs and outputs for the SBoM of JavaScript application bundles in sliding windows. $W$ represents a single sliding window and $X$ represents an input sequence. $\hat{X}$ and $\overline{X}$ represent a clone and non-clone script.
Each window has $win\_size$ number of tokens with a class token for segmentation classification purpose. 
$N$ windows construct a sequence input with size $seq\_len$ such that $X^i \in \mathbb{Z}^{win\_size + 1}$ and $X \in \mathbb{Z}^{seq\_len + N}$.
The outputs $<\mathds{1}_{seg}, y_{cls}>$ correspond to the boundary indicator and class name described in the previous section.
$\{-1, 1\}_{ccr}$ denotes the similarity score calculated from cosine similarity, where one indicates the exact match and vice versa.
In addition, each sliding window is incremented by the size of $step$.
During training, we apply the greedy sliding window approach where we set the step size to be the same as the sequence size.
During the inference phase, we slide one token each step ($step=1$) to accumulate the prediction outputs for enhanced precision.
%\yuan{not sure if i get this, window size is a fixed value or length of the testing sequence? why setting 1 is helpful?}\comment{fixed, make more clear}.
% The second approach called the full sliding window approach can be used to improve the inference accuracy.
% We set the step size to 1, and accumulate the prediction results for each token.
% Then we can set a threshold to predict the location of the boundary and rank the accumulated class names and clone target names.
% The reason under the hood is that if the step size is 1, one window being predicted as true means the boundary is at the end of the window.
% Therefore, we can skip the whole window to predict the boundary for the next window.

\subsection{CoE Model Architecture} \label{sec:coe}
The CoE model represents a novel approach to train a unified, multi-task model for a series of sequential tasks, akin to an assembly line. In general, a CoE model integrates M backbone neural networks, denoted as $B_1, B_2, ..., B_M$, and N expert neural networks, represented as $E_1, E_2, ..., E_N$. The output of the nth expert model can be formulated as the following, where $\circ$ stands for the function composition:
\begin{equation}
    B_1 \circ B_2 \circ ... \circ B_m \circ E_n
\end{equation}
In this architecture, each expert benefits from the information processed by the preceding task, creating a coherent data flow.

\noindent \textbf{Backbone Models.} The foundational backbone model, $B_1$, is a pre-trained BERT encoder, such as SentenceBERT~\cite{reimers2019sentence} or GraphCodeBERT~\cite{guo_graphcodebert_2021}. These models, without the use of causal attention masking~\cite{devlin2018bert}, enhance the semantic understanding of code in a bidirectional context. The latter backbone models employ a Transformer encoder structure~\cite{vaswani_attention_2017}. We stack untrained backbone and expert models on top of a pre-trained encoder to simulate the use of pre-trained models for downstream tasks. Similar to the BERT-style architecture, the $B_2$ encoder structure does not use any causal masking. Instead, to prevent data overflow to adjacent segments, we apply a novel segmentation masking in this layer (see Section~\ref{sec:seg_masking}). 

\noindent \textbf{Expert Models.} The expert model for code segmentation is constructed with a linear output layer, parameterized by the weight matrices $\textbf{W}_{seg} \in \mathbb{R}^{d \times 2}$. The expert model for code classification is a linear output layer, parameterized by the $\textbf{W}_{cls} \in \mathbb{R}^{d \times c}$. The code clone retrieval expert model takes two vectors with size $\mathbb{R}^{d}$ to facilitate further cosine similarity matching. 

\noindent \textbf{Main Algorithm.} As CoE is a multi-level structure, we can incorporate previous windows for extended input length. The input for the code segmentation layer is sourced from the first \textit{cls} token of the encoder output. Meanwhile, the inputs for the code classification layer and the code clone retrieval layer are derived from each \textit{cls} token from the $B_2$ layer. The process can be summarized in Algorithm~\ref{algorithm}.
\begin{algorithm}
\caption{CoE is a chain of deep neural networks for code segmentation, code classification, and code clone retrieval} \label{algorithm}
\begin{algorithmic}[1]
\Require $X, \tilde{X}$
\State Initialize $B_1$ as a $SentBERT$ layer, $B_2$ as a $TransEnc$ layer
\State Initialize $E_{seg}$ as a composition of a $TransEnc$ layer and a linear transformation layer parameterized by $\textbf{W}_{seg}$
\State Initialize $E_{cls}$ as a linear transformation layer parameterized by $\textbf{W}_{cls}, \textbf{d}$
\State Initialize $E_{ccr}$ as a cosine similarity function
\State $\textbf{X} \gets Embedding(X)$
\For{each sliding window $\textbf{X}^i$ in $\textbf{X}$}
    \State $\textbf{h}_1^i \gets SentBERT (\textbf{X}^i)$ \Comment{$\textbf{h}_1^i \in \mathbb{R}^{(win\_size+1) \times d}$}
\EndFor
\State $\textbf{h}_2 \gets \textbf{h}_1^1[0] \: || \: \textbf{h}_1^2[0] \: || \: ... \: || \: \textbf{h}_1^N[0]$ \Comment{$\textbf{h}_2 \in \mathbb{R}^{N \times d}$}
\State $\textbf{o}_{seg}^i \gets \textbf{W}_{seg} \cdot TransEnc(\textbf{h}_2)^i $ \Comment{$\textbf{o}_{seg}^i \in \mathbb{R}^{2}$}
\State $seg\_mask \gets f_{binary}(\textbf{o}_{seg})$
\State $\textbf{h}_3 \gets TransEnc (\textbf{h}_2, seg\_mask)$
\State $\textbf{o}_{cls}^i \gets \textbf{W}_{cls} \cdot \textbf{h}_3^i$
\Comment{$\textbf{o}_{cls}^i \in \mathbb{R}^{c}$}
\State $o_{ccr}^i \gets cos\_sim(\textbf{h}_3^i, \tilde{\textbf{h}}_3^i)$ \Comment{$o_{ccr}^i \in [-1, 1]$}
\Statex \hspace*{-\algorithmicindent}\textbf{Return:} $\textbf{o}_{seg}, \textbf{o}_{cls}, o_{ccr}$
\end{algorithmic}
\end{algorithm}

Algorithm~\ref{algorithm} explains the data throughput across all the backbone models and expert models.
In the actual implementation of this algorithm, line 5 is replaced with a reshape operation from shape $batch \times (seq\_len+N) \times d$ to $(batch \times N) \times (win\_size+1) \times d$.
After the reshape operation, the first token of each window is grouped into $\textbf{h}_2$, replacing the concatenation operation on line 9 of Algorithm~\ref{algorithm}.
During the training phase, the model processes pairs of code snippets for code clone similarity.
During the inference phase, the model only requires one input, whereas, those code snippets in the database $\mathcal{D}$ are embedded for clone retrieval.

In essence, the CoE model offers a coherent and integrated solution for complex, multi-stage tasks in software analysis, leveraging the strengths of each component to deliver multiple outputs for all tasks.

\subsection{Segmentation Masking} \label{sec:seg_masking}
\begin{figure}[htbp]
    \centering
    \includegraphics[width=0.6\columnwidth]{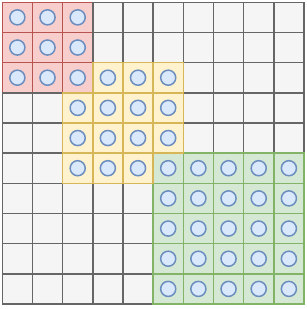}
    \caption{The figure illustrates the segmentation masking, where different color blocks are activated attentions and grey areas are masked, such that tokens can only be attended to each other within the same segment.}
    \label{fig:masking}
\end{figure}

Lines 11 and 12 of Algorithm~\ref{algorithm} detail the generation and application of segmentation masking.
Figure~\ref{fig:masking} explains the concept of segmentation masking\yuan{\sout{where do we mention this segmentation masking in Algorithm 1? }}.
In the illustrated figure, different-colored blocks represent segments, where attention activation is permitted.
This indicates areas of the input data that the model is allowed to process in relation to each other.
The grey areas, on the other hand, are masked, preventing the model from attending to these regions.
This ensures that each token in the input data is only compared with other tokens within the same segment, effectively isolating distinct segments.
In addition, the masking blocks interact with each other at the boundary windows, as such windows can contain code tokens from adjacent segments.
Remember each token represents a window in this context.

%\yuan{i did not get, this is a new design for what scenarios? can you use clone retrieval or code classification as example?}\comment{explained}
This attention restriction brought about by segmentation masking leads to more meaningful interaction among data points within the same context, avoiding irrelevant or noisy signals from outside the segment. As a result, the model can perform more nuanced and context-aware analysis, such as code classification and data compression for clone retrieval. For example, in Figure~\ref{fig:masking}, each color black represents a segment predicted by expert model $E_{seg}$. By applying the segment masking, the latter backbone and expert models will only use the tokens from the same segment for classification and embedding. 
In addition, we choose not to implement causal masking for segment masking as it is a common practice in embedding models.
The bidirectional attention allows for an even distribution of contextual information to each token~\cite{reimers2019sentence}.
Consequently, the segmentation masking method facilitates the application of a window-wise loss, as detailed in Section~\ref{sec:loss}.

\subsection{The joint losses of three experts} \label{sec:loss}

The joint loss function for the CoE model, as it pertains to the tasks of SBoM, is listed in Equation~\ref{eq:joint_loss}. This function integrates losses from three distinct tasks: code segmentation, code classification, and code clone retrieval. Each component of the loss function addresses a specific aspect of the overall task. The superscript $i$ is the window index, $N$ is the number of windows, and $B$ is the number of samples per batch.

\begin{equation} \label{eq:joint_loss}
\begin{aligned}
\mathcal{L}^i_{CoE} &= \mathcal{L}^i_{seg} + \mathcal{L}^i_{cls} + \mathcal{L}^i_{ccr} \\
&= CE(\textbf{o}^i_{seg}, y^i_{seg}) + CE(\textbf{o}^i_{cls},  y^i_{cls}) + \{-1, 0, 1\} \cdot \textbf{o}^i_{ccr} \\
\mathcal{L}_{CoE} &= \frac{1}{BN} \sum_B \sum_N^{i=1} \mathcal{L}^i_{CoE}
\end{aligned}
\end{equation}

\noindent \textbf{Code Segmentation Loss ($\mathcal{L}_{seg}$)}: This component, represented as $CE(\textbf{o}^i_{seg}, y^i_{seg})$, employs cross-entropy (CE) to measure the difference between the predicted window ($\textbf{o}^i_{seg}$) and the label of this window ($y^i_{seg}$) in a JavaScript file. It is essentially a binary classification problem, which quantifies how well the model can segment a file into logically coherent parts with distinct semantics.

\noindent \textbf{Code Classification Loss ($\mathcal{L}_{cls}$)}: Given as $CE(\textbf{o}^i_{cls}, y^i_{cls})$, this part of the loss function also uses cross-entropy to assess the accuracy of the model in classifying the identified code segments. Here, $y^i_{cls}$ represents the actual class label of the window, and $\textbf{o}_{cls}$ denotes the predicted probabilities by the model.

\noindent \textbf{Code Clone Retrieval Loss ($\mathcal{L}_{ccr}$)}: This loss is represented as $\{-1, 0, 1\} \cdot \textbf{o}^i_{ccr}$. 
The cosine similarity is computed between a query vector $\mathbf{h}$ and a potential clone $\tilde{\mathbf{h}}$.
For the code classification task, if the label is predicted not to be in the expected dataset $\mathcal{D}$, we multiply the cosine similarity loss by 0 to stop the gradients for such segments.
As deep neural networks are optimized by gradient descent, we multiply the cosine similarity score by $-1$ if the pair is a clone.
Contrastively, we multiply by $1$ if the pair is not a clone.

This joint loss function per window, $\mathcal{L}_{CoE}$, effectively combines these three aspects to train the CoE model comprehensively.
The final loss is calculated as the mean of the individual losses across all windows within a single batch.
It is important to highlight that in the CoE model, each component of the joint loss function, as described in Equation~\ref{eq:joint_loss}, is not scaled by any additional parameters. This design choice ensures that the losses from the code segmentation task ($\mathcal{L}{cs}$), the code classification task ($\mathcal{L}{cc}$), and the code clone retrieval task ($\mathcal{L}_{ccr}$) contribute equally during back-propagation to both the backbone models and the expert models. 

\subsection{BPE Tokenization}
\yuan{could this be part of the implementation detail? in Section IV}
We tokenize each file by Byte-Pair Encoding (BPE)~\cite{shibata1999byte}. This approach mitigates out-of-vocabulary issues and the impact of JavaScript minification, where spaces or indents are typically excluded. BPE offers the advantage of not requiring additional JavaScript parsing, enabling raw JavaScript files to be directly inputted into neural networks, thus enhancing serialization speed. It is noteworthy to mention certain caveats encountered during data generation, such as the exclusion of the last segment and the importance of spaces and line breaks during tokenization, as they might hint at segment boundaries in a bundle.

\section{Experiment Setup}

\subsection{Data collection and preparation}
To assess the generalizability of our methodologies, we retrieve real-world JavaScript application bundles from NPM Registry\footnote{\href{https://www.npmjs.com/}{NPM Registry: a vast online database of open-source JavaScript packages and libraries}}.
Given the vast number of available JavaScript packages, we randomly download 4000, 500, and 500 packages for training, validation, and testing, respectively. We discovered that the 4000 JavaScript application bundles downloaded are sufficient to create an adequate number of data segments and windows for training purposes (see Table~\ref{table:dataset}). Each application bundle is accompanied by a source mapping file that labels the location of the boundary and the source name, which indicates the type of the segment.
In addition, the source mapping file has a ``sourceContents'' field which links each segment to its original source code before optimization and compression. We aggregate all the sources as the embedding corpus for the code clone retrieval task.
Typically, this source mapping file is not published in JavaScript deployments.
We utilize the source mapping files only to produce the ground truth for training and evaluation purposes.

%\yuan{for the retrieval task, i guess you need a search target corpus, what is that?}\comment{explained above}

Table~\ref{table:dataset} lists the detailed statistics of the downloaded and generated datasets.
By default, the length of each sliding window is 512.
The number of samples is the same across all three tasks.
In addition, as discussed in Section~\ref{sec:sw} for the sliding windows approach, each window can include zero or more boundaries.
To categorize these windows, we determine their labels based on the predominant class type, which is identified by the largest number of tokens present within the window.

\begin{table}[h]
\centering
\caption{The statistics for the dataset, including the number of source files, the average number of tokens in each file, the total number of segments, and the number of samples for 512 tokens long sliding windows at each phase. A bunlde file consists of multiple files and one segment can be divided into multiple windows, where sliding windows are the inputs for the neural networks.\yuan{\sout{SW means what?why worth to be mentioned? what's the differences between number of segments and number of samples for 512 sliding windows? A bit hard to link them to the requirement of CoE}} }
\vspace{0.3cm}
\begin{tabular}{lrrr}
\hline
                    & \multicolumn{1}{l}{Training} & \multicolumn{1}{l}{Validation} & \multicolumn{1}{l}{Test} \\ \hline
Number of Files               & 4,000                        & 500                            & 500                      \\
Tokens per File (Avg)        & 82,068                       & 71,202                         & 66,723                   \\
Number of Segments & 608,790 & 75,082 & 72,566 \\ 
Number of Windows (512)  & 301,487                      & 30,850                         & 66,079                   \\ \hline
% Code Classification & 158,815                      & 19,819                         & 20,015                   \\
% Code clone retrieval   & 608,790                      & 75,082                         & 72,566                   \\
\end{tabular}
\label{table:dataset}
\end{table}

Next, we explain the process of preparing the dataset for evaluating individual tasks.

\subsubsection{Code Segmentation}
% For the code segmentation task, we label each token based on the presence of a segment following that token. 
For the code segmentation task, we label each window based on the presence of the boundary within the window.
The accompanying source map file with each package allows us to determine segment boundaries by decoding the Base64 VLQ.
Then, we slice each script into windows with fixed size as the method discussed in Section~\ref{sec:sw}.
Since the number of windows with a boundary is significantly less than the number of windows without a boundary, we downsample false labels to match the number of true labels.
For the test set, we include all the windows without downsampling to match the real-life scenario.
In total, there are 301,000, 31,000, and 66,000 samples for training, validation, and testing\yuan{why we have a different set for training and evaluating of one task? shouldn't CoE be end-to-end, which means the same data will be used to train and complete three tasks?}\comment{different bundles have different compositions of segments, even same source can be compressed into a different set of tokens. Also, the embedding model doesn't need to see all the sources.}.

\subsubsection{Code Classification}
In the source map file, each segment is labeled with its source name.
We classify the source names into three categories: \textbf{bootstrap segments}, \textbf{application files}, and \textbf{third-party libraries}.
Bootstrap segments are generated by JavaScript bundlers to bundle the imported files.
Application files are local native code snippets that specifically belongs to this application during the bundling process. 
Third-party libraries are imported files from the NPM repository. The labels are generated from the source name of each segment. In the future, we can adjust the labels set to conform with the SBoM format, such as the type file in CycloneDX~\cite{cyclonedx}.
%}\yuan{how they are labeled? or you pull them from different sources?}\comment{fixed}

\subsubsection{File-level Code Clone Retrieval}
As the smallest scope of a JavaScript segment is the file level, we aim to retrieve those segments that are labeled with \textbf{third-party libraries} from the existing NPM repositories.
During the training phase, we pair a clone or a non-clone with the segment generated by the code segmentation task and for cosine similarity (see Section~\ref{sec:sw}).
For the validation and test phases, we make a pair-wise comparison with the files in a repository and rank them based on the highest cosine similarities.

\subsection{Baselines}
\yuan{\sout{i think we should highlight that there is no solution for this pipeline, but many ways to implement the pipeline we proposed for SBoM for javascript, thus we pick baselines based on their capability to completing one or more tasks in the proposed pipeline.}} 
As discussed in the introduction, there is no single solution for this pipeline. By dividing the SBoM generation into three sub-tasks, we can pick baselines in each area.
Some of the baselines are used for more than one experiment, such as Transformer, Longformer, and SentBERT.
Since we use SentBERT as the backbone model and the two-level architecture similar to CATS, we can compare the results for ablation studies.

%\yuan{what are the models you considered as baselines for each task? are the first four general models used in three tasks and the final two for classification and retrieval?}\comment{fixed}

% \subsubsection{Code Segmentation and Classification}
\noindent \textbf{Transformer}~\cite{vaswani_attention_2017} model, with its self-attention mechanism, has demonstrated impressive performance across various NLP tasks. We use it as a baseline for all the tasks.

\noindent \textbf{CATS}~\cite{glavas_two-level_2020}, a two-Level Transformer, is the state-of-the-art approach for text segmentation challenges. Since we concatenate the segmentation expert with backbone model $B_1$, the segmentation architecture is actually a one-level architecture, which is not the same as CATS.

\noindent \textbf{LongFormer}~\cite{beltagy2020longformer} is a variant of the BERT architecture to support long sequences.The core novelty of the Longformer lies in its attention mechanism, which combines both local and global attention. We aim to test the performance of long sequences segmentation and embedding.

\noindent \textbf{SentBERT}~\cite{reimers2019sentence} is a light weight pre-trained BERT for sentence embedding. SentBERT is well-suited for segmentation, classification and embedding.

\noindent \textbf{Token-based Code Clone Retrieval} method directly applies the cosine similarity static function to the numeric code tokens as a baseline to indicate the similarity of the input and output code snippets.

\noindent \textbf{Gzip + KNN}~\cite{jiang2023low} is a novel approach for text classification with only a compression method and KNN algorithm. Such a method can support sequences with any length and search the target among a very large dataset, which is suitable for the clone retrieval experiment setup.

% The \textbf{GraphCodeBERT}~\cite{guo_graphcodebert_2021} model is the state-of-the-art model in code processing tasks. However, the number of tokens of GraphCodeBERT is limited to 512.

\subsection{Evaluation Metrics}
\subsubsection{Code Segmentation}
The performance metrics for the code segmentation task include \textbf{accuracy}, \textbf{precision}, \textbf{recall}, Area Under the Receiver Operating Characteristic Curve (\textbf{AUROC}), processing speed in tokens per second (\textbf{Tokens/s}), and \textbf{the number of model parameters}. The AUROC is a critical metric for assessing the quality of a model's classifications, especially in cases of imbalanced datasets. Unlike accuracy, which can be misleadingly high when one class significantly outnumbers the other, the AUROC accounts for both the true positive rate (sensitivity) and the false positive rate across different decision thresholds. It provides a robust, threshold-agnostic measure of a model's ability to discriminate between positive and negative classes. A value of \(1\) signifies perfect classification, while a score of \(0.5\) suggests that the model is performing no better than random chance. For tasks such as code segmentation, where the segments of interest might be infrequent relative to the overall codebase, AUROC serves as a more informative metric for model evaluation, capturing the model's capability to identify these rare segments correctly.

\subsubsection{Code Classification}
For the code classification task, we employ two key metrics to assess the model performance and applicability:
\textbf{Accuracy} gauges the model's capability to correctly identify code clones, providing a straightforward measure of its classification effectiveness, by comparing correct predictions against the total predictions.
As discussed, \textbf{AUROC} is vital in a multi-class classification experiment.
It should be noted that the processing speed is the same for the code segmentation task.

\subsubsection{Code Clone Retrieval}
For the code clone retrieval task, \textbf{accuracy} calculates the correct number of identified clones. As mentioned in the data pipeline section, we calculate the similarity scores across each pair of files. 
The \textbf{time} metrics evaluate the temporal cost of the similarity matching. For those methods that can encode the file into a fixed-length vector, we can apply Approximate Nearest Neighbours~\cite{8594636} to accelerate the search speed.

% \subsection{Training Details}
% We train the model on four RTX 6000 GPUs.
% The input length for the CoE model is 2044 with 4 windows and 511 tokens within each window.

% \begin{table}[hb]
% \begin{tabular}{l|llll}
% \hline
% X          & win\_size & N   & seq\_len &       \\
%            & 511       & 4   & 2044     &       \\ \hline
% $B_2$       & layers    & d   & heads    & d\_ff \\
%            & 4         & 768 & 12       & 2048  \\ \hline
% $E_{seg}$ & layers    & d   & heads    & d\_ff \\
%            & 2         & 768 & 12       & 2048  \\ \hline

% \end{tabular}
% \end{table}

%\subsection{CoE Model Setup} \yuan{why this is a part of baseline?}\yuan{why we only talk about base model not expert model setup?}\yuan{setup is usually after evaluation metrics}\comment{fixed}
\subsection{Implementation}
%Our model is fine-tuned with key hyperparameters to enhance performance. 

We set the feature dimension to 768, a standard for BERT-based models, ensuring efficient vector representation. The backbone, SentBERT ($B_1$), with its 33 million parameters, provides a strong foundation for sentiment analysis. We augment this with one linear layer for the Segmentation Expert ($E_{seg}$).
% , each with a feed-forward dimension of 3072 and 12 attention heads, which adeptly captures sequence relationships.
Additionally, our model includes 4 layers of Transformer decoders with Segmentation Masking (backbone $B_2$), also configured with 3,072 feed-forward dimensions and 12 attention heads for the classification and embedding tasks. This configuration strikes an optimal balance between complexity and computational efficiency.
The Classification Expert ($E_{cls}$) is a linear layer mapping from 768 hidden dimensions to the number of classes.
The embedding output is the same output of the $B_2$.
For the sliding window configuration, CoE takes 4 windows with 512 tokens each.
As a result, the input length of CoE is 2,048 tokens without further notice.

\yuan{\sout{usually, (if have space), we can mention which computer is used for running experiments. For instance, Our computational infrastructure used for the experiments consists of 6x NVIDIA RTX A6000 GPUs on an AMD EPYC Server with 128 CPU cores. }} We trained and evaluated our models on a server with 4 NVIDIA RTX A6000 GPUs and 64 AMD EPYC CPU cores. The training time varies from 12 hours (2 sequences per batch per GPU for Longformer) to 4 hours (8 sequences per batch per GPU for SentBERT).

\begin{table*}[ht]
\centering
\caption{Performance of code segmentation models on real-world JavaScript application bundles from NPM.}\label{tab:seg}
\vspace{0.2cm}
\begin{tabular}{lcccccc}
\hline
Model         & Accuracy         & Precision     & Recall        & AUROC        & Tokens/s        & Parameters \\ \hline
Transformer   & 87.39\%          & 0.89          & 0.85          & 0.95          & 19,660          & 38 M       \\
              & 88.51\%          & \textbf{0.92} & 0.84          & 0.96          & 4,515           & 124 M      \\
CATS          & 88.36\%          & 0.89          & 0.88          & 0.96          & 24,576 & 38 M       \\
              & 87.74\%          & 0.91          & 0.84          & 0.95          & 5,698           & 124 M      \\
Longformer   
% & 86.16\%          & 0.90          & 0.82          & 0.94          & 22,282          & 38 M       \\
              & 87.15\%          & 0.84          & \textbf{0.91} & 0.95          & 4,735           & 148 M      \\
SentBERT      &89.08\%	   &0.77	&0.86	&0.95	&\textbf{27,716}	&33 M	\\
% GraphCodeBERT & 91.58\% & \textbf{0.92} & \textbf{0.91} & \textbf{0.97} & 3,471           & 124 M      \\
\hline
CoE & \textbf{90.44\%} & \textbf{0.92} & \textbf{0.91} & \textbf{0.97} & 25,461           & 38 M      \\ \hline
\end{tabular}%
\end{table*}

\section{Results} \label{sec:experiment}
We conduct thorough evaluations focusing on three primary tasks: code segmentation, code classification, and code clone retrieval.\yuan{\sout{ feel we already mention the below setup in the previous paragraph? if so, can be removed, unless your publicly available JS package means something else}} Our methodologies are benchmarked based on two criteria: overall accuracy and time performance metrics.

\subsection{Code Segmentation Task}

Table \ref{tab:seg} presents the evaluation of different code segmentation models using a dataset comprising of JavaScript packages downloaded from NPM.  The models evaluated are Transformer, CATS, Longformer, SentBERT, and CoE, each with varying configurations.

The CoE model achieves the highest overall performance with an accuracy of \(91.92\%\), precision and recall rates at \(0.92\) and \(0.91\) respectively, and an AUROC of \(0.97\). It also features a sizable parameter count of 38 million with a processing speed of 25,461 Tokens/s. These results underscore the effectiveness of incorporating a shared backbone for all downstream tasks, even though the model comes with higher computational costs.

Transformer-based models, in their varying configurations, also deliver robust performances, but are underperformed by specialized models, such as CoE. 
The CATS model, utilizing a two-level architecture for code segmentation, has high processing speed as other models.
Similarly, Longformer offers reasonable performance metrics, but falls short in terms of accuracy and AUROC compared to CoE. These findings suggest that while transformer-based models are highly versatile, specialized architectures tailored for code segmentation offer superior performance at the cost of computational efficiency.
The SentBERT model, on the other hand, demonstrates the highest processing speed, capable of processing \(27,716\) tokens/s with 33 million parameters.
In general, the processing speed scales to the number of parameters of the neural networks.
However, its accuracy, precision, and recall rates are slightly lower than those of CoE, highlighting a trade-off between computational efficiency and model performance.

\begin{figure}[ht]
    \centering
    \includegraphics[width=\columnwidth]{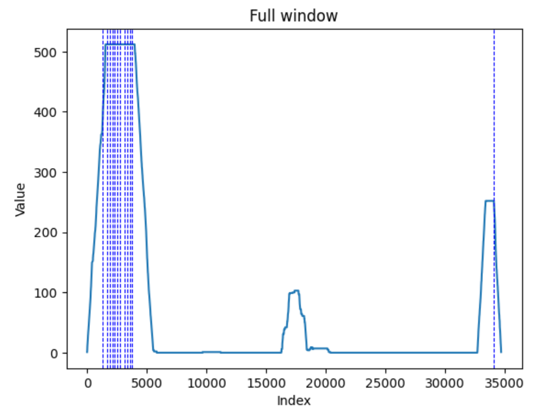}
    \caption{Actual boundaries (vertical dotted lines) vs accumulated predictions (the solid line) by one step sliding window for a bundle file with 35,000 tokens.}
    \label{fig:full_sw}
\end{figure}

Figure \ref{fig:full_sw} illustrates the application of the full sliding window approach to detect JavaScript package boundaries in a script with 35,000 tokens, that is the horizontal axis. The vertical axis, labeled "Value" denotes the cumulative prediction counts for each token, with potential values ranging from 0 up to a maximum of 512 — this maximum arises from the window size used.
The sliding window has a fixed size of 512 tokens and moves with a step size of 1 token. At each step, the model makes a binary prediction (either 0 or 1) indicating the absence or presence of a package boundary. The solid line on the graph represents the summed predictions for each token, effectively indicating how many times a particular token has been identified as a boundary across all overlapping windows.
Distinct peaks in the blue line represent high consensus among overlapping windows that a particular token is at or near a package boundary. The vertical dotted lines, distributed across the figure, indicate the true boundaries of the JavaScript packages. These serve as a ground truth against which the model's predictions, illustrated by the solid blue line, can be evaluated. The close alignment between peaks in the prediction and the dotted lines would indicate accurate boundary detection by the model. Conversely, gaps or discrepancies would suggest areas where the model might have faced challenges or made errors in boundary identification.

\subsection{Code Classification}

Table \ref{table:cls} lists the results for the Code Classification task.
The CoE model, integrating multiple expert models to leverage learned knowledge across different tasks, records an accuracy of 77.34\% and the highest AUROC of 0.91 among the compared models. This underscores CoE's advanced capability in distinguishing between categories with a higher degree of certainty, attributed to its comprehensive learning strategy that benefits from task interdependencies.
\begin{table}[h]
\centering
\caption{Code classification results.} \label{table:cls}
\vspace{0.3cm}
\begin{tabular}{lll}
\hline
Model       & Accuracy & AUROC \\ \hline
Transformer & 72.43\%  & 0.88   \\
SentBERT    & \textbf{78.57\%}  & 0.86   \\\hline
CoE         & 77.34\%  & \textbf{0.91}   \\ \hline
\end{tabular}
\end{table}
These results underscore the varying strengths of each model, with CoE's balanced performance highlighting the advantage of its multi-task learning approach in complex code classification scenarios.

\subsection{Code Clone Retrieval}

We compared the performance of four different algorithms for the task of clone retrieval within a dataset of 100,000 files. 
The Naive method served as our baseline, and as expected, it showed zero percent accuracy.
The Gzip + KNN approach demonstrated a decent accuracy of 70.88\% but was considerably slower with an average time of 2.66 seconds per file.
Longformer did not yield any accurate results, and its processing speed was relatively slow at 3.7 files/s. SentBERT also has low accuracy for code clone search. This calls into question its practicality for dynamic and ever-evolving code repositories. In contrast, the CoE model yields the highest accuracy of 73.77\% and supports ANN.
The disparity in the time and accuracy metrics across the algorithms highlights the computational-efficiency-versus-accuracy trade-off inherent in clone retrieval tasks.

\begin{table}[h]
\centering
\caption{Performance of code clone retrieval models on segmented scripts.}
\vspace{0.3cm}
\begin{tabular}{lcll}
\hline
Model                & Accuracy & Time  (Average)                   \\ \hline
Token-based                & 0.00\%                   & Supports ANN                      \\
Gzip + KNN           & 70.88\%                  & 2.66 s/file                       \\
LongFormer-pairwise*       & 0.00\%                   & 3.7 files/s                       \\
SentBERT          & 0.17\%                   & Supports ANN                      \\ \hline
CoE                  & \textbf{73.77\% }                 & Supports ANN                       \\ \hline
\multicolumn{3}{l}{\multirow{2}{*}{\begin{tabular}[c]{@{}l@{}}*LongFormer processes a unified window \\ with 2048 tokens.\end{tabular}}} \\
\multicolumn{3}{l}{}    
\end{tabular}\label{table:ccr}
\end{table}

\section{Discussion}
\subsection{Strengths}
\noindent \textbf{CoE performs better than individual single expert.}
The CoE model demonstrates better performance compared to individual expert models across a wide array of tasks.
CoE enables each subsequent expert in the sequence to operate on a more informed basis. This not only enhances decision-making capabilities but also significantly boosts the model's overall accuracy and efficiency.
Moreover, the CoE model's architecture promotes robust performance by mitigating the likelihood of errors through cross-validation among experts. This leads to a more reliable output compared to that of single experts, which may not fully capture the breadth of data or adapt as readily to varied task requirements.

\noindent \textbf{CoE requires less training time and inference time.}
The CoE model achieves a more efficient training process by incorporating shared learning, sequential knowledge transfer, focused training efforts, optimized back-propagation, and parallel processing. These features collectively reduce the time required to train the model across multiple tasks, making CoE an attractive solution for multi-task learning scenarios where time and computational resources are critical considerations.

% \noindent \textbf{Robust segmentation.} A vital aspect of our solution is the capability to segment the input effectively. This is evidenced by our model's high recall and Area Under the Receiver Operating Characteristic (AUROC) values, which signify both the sensitivity of our model and its ability to discriminate between positive and negative classes.
    
% \noindent \textbf{Semantic Encoding.} For semantic encoding, it is crucial to maintain the integrity of the information. Our approach ensures that minor alterations or nuances in the data don't adversely affect the encoding outcome. This is achieved by implementing mechanisms to rule out the impacts of these trivial modifications.
    
% \noindent \textbf{Generalizability.} Beyond the initial training data, it's imperative for our model to exhibit efficacy on unseen or out-of-sample data. Our solution is architected to ensure that the model remains adaptable and competent in incorporating out-of-sample classifications, reinforcing its versatility.
    
% \noindent \textbf{Performance trade-off.} In any computational solution, there often exists a trade-off between accuracy and speed. Our model has been fine-tuned to strike an optimal balance, ensuring high performance without compromising on the quality of the results.

\noindent \textbf{Maximum sequence length.}
Utilizing a two-level Transformer architecture in our model significantly expands the potential maximum sequence length that can be effectively processed.
In theory, the architecture allows for a sequence length of up to $262,144$, $512 \times 512$, where each level of the Transformer can handle up to $512$ tokens. This extended length not only accommodates longer individual sequences but also enables the model to manage multiple sequences in parallel, which is a common scenario in many real-world applications. The choice of 512 as the optimal input size for both the SentBERT and Transformer Encoder models is based on empirical evidence that suggests this length strikes an optimal balance between computational efficiency and model performance. At this size, the models can capture the necessary contextual information without being overwhelmed by excessive data, leading to improved accuracy and faster training times.
In practice, the GPU memory can limit the maximum sequence length.

% \begin{table}[]
% \begin{tabular}{lllll|lllll}
% \hline
% \multicolumn{5}{c|}{Train on}   & \multicolumn{5}{c}{Test on}      \\ \hline
% N & win\_size & seg & cls & ccr & N  & win\_size & seg & cls & ccr \\
% 4 & 512       &     &     &     & 80 & 512       &     &     &     \\
% 8 & 256       &     &     &     & 80 & 256       &     &     &     \\ \hline
% \end{tabular}
% \end{table}

\noindent \textbf{CoE is an end-to-end solution.}
In summation, the CoE model has been designed and empirically validated as an end-to-end solution for the SBoM in the domain of JavaScript application bundles. The CoE framework adeptly manages the intricate processes of code segmentation, classification, and clone retrieval through a unified, streamlined approach, ensuring that each stage of the analysis is both informed by and contributes to the stages that follow. By facilitating a deep, neural network-driven analysis of JavaScript bundles without the need for intermediary steps or external dependencies metadata, CoE embodies a comprehensive solution that spans the full breadth of the SBoM creation process. The end-to-end nature of CoE is a testament to its ability to handle the complete lifecycle of SBoM tasks, delivering a solution that is not only efficient and scalable but also significantly reduces the complexity and time required for such in-depth security analyses.

\subsection{Model Variants and Ablation Studies}

\noindent \textbf{Not sensitive to loss coefficients.} In the context of Multi-Task Learning (MTL), balancing the contributions of different tasks in the joint loss function is a complex issue that has been the subject of extensive research~\cite{chen2018gradnorm}.
Despite this, the precise coefficients (\(\alpha\), \(\beta\), and \(\gamma\)) used in the joint loss function, as depicted in the equation below, have been found to have minimal impact on the final performance of the model, provided that the performance standard deviation remains within \(\pm 0.1\).

\begin{equation}
\mathcal{L}_{\text{CoE}} = \alpha \mathcal{L}_{\text{seg}} + \beta \mathcal{L}_{\text{cls}} + \gamma \mathcal{L}_{\text{ccr}}
\end{equation}

It suggests that the choice of coefficients does not need to be overly fine-tuned. Instead, a more robust approach can be taken where the coefficients are set to values that intuitively balance the tasks, without the need for an exhaustive search for the optimal values. This simplification can greatly reduce the complexity of the model's fine-tuning process, especially in scenarios where the tasks are of similar difficulty or when prior knowledge about the relative importance of the tasks is limited. Moreover, this finding aligns with the understanding that in MTL~\cite{crawshaw2021slaw}, the structure and the relationship between tasks are often more critical than the exact loss weights. By ensuring that the tasks are well-balanced in terms of data availability, complexity, and relevance, and by using a well-designed joint architecture that can effectively share knowledge between tasks, the model can achieve consistent performance across different tasks without the need for meticulous tuning of the loss coefficients.

% \noindent \textbf{Linear Transformations are good enough for Expert models.}
% We tried different model structures for expert models. We find that fully connected layers are good enough for decent performance.

\noindent \textbf{Sliding windows configurations.}
To find the best sliding window configurations, we record validation performance three times for each configuration.
According to Table~\ref{table:val_seg_len}, the performance of the code segmentation task does not always scale with the window size.
\begin{table}[h]
\caption{Validation performance with different sliding window configurations.}  \label{table:val_seg_len}
\vspace{0.2cm}
\begin{tabular}{ll|lll}
\hline \hline
N & win\_size & seg (AUROC) & cls (AUROC) & ccr (Acc) \\ \hline
4 & 512       & 96.65\% $\pm$ 0.0  & 84.48\%  $\pm$ 0.0      & 73.77\%  $\pm$ 0.2      \\
2 & 512       & 96.55 $\pm$ 0.1 & 81.72\% $\pm$ 0.2          &  67.53\%  $\pm$ 0.8         \\
2 & 256       & 97.04\% $\pm$ 0.2        & 86.39\% $\pm$ 0.   & 74.49\%  $\pm$ 1.1    \\
4 & 256       & 97.18\% $\pm$ 0.2  & 87.89\% $\pm$ 0.1   &   74.82\%  $\pm$ 0.4        \\
4 & 128       & 97.75\%  $\pm$ 0.0   & 87.46\% $\pm$ 0.0     & 74.23\%  $\pm$ 0.5 \\ 
8 & 256       & 97.55\%  $\pm$ 0.0   & 87.05\% $\pm$ 0.0     & 74.35\%  $\pm$ 0.3   \\
8 & 128       & \textbf{97.92\%  $\pm$ 0.0}   & \textbf{89.52\% $\pm$ 0.1}     & \textbf{75.58\%  $\pm$ 0.2}  \\ \hline \hline
\end{tabular}
\end{table}
The sliding windows configuration is fine-tuned for optimal performance in code segmentation with a window size of 256 tokens. This configuration effectively balances the need for contextual understanding with computational efficiency, making it ideal for parsing code into meaningful segments. For broader analysis tasks such as code classification and code clone retrieval, performance scales positively with the total sequence length ($N \times win\_size$). These tasks benefit from analyzing longer sequences to identify cross-functional patterns and clones, as a larger sequence length provides the model with a more comprehensive view of the codebase.
In addition, the performance of the code clone retrieval task is close for different configurations, but showing high variances.

\noindent \textbf{Segmentation masking.}
Table~\ref{table:val_seg_masking} lists the AUROC performance of the validation set for the code classification task.
The `one seg' column denotes a classification task on one segment only.
The 93.76\% AUROC of this task indicates that classifying on perfectly segmented blocks can significantly improve the performance.
Without the segment masking, the model demonstrates poor performance on classification tasks.
In contrast, by restricting the data information within a segment, the performance of CoE can approach the `one seg' task.

\begin{table}[h]
\centering
\caption{Model performance of validation dataset with and without segmentation masking, and excluding segmentation as a prerequisite task.}  \label{table:val_seg_masking}
\vspace{0.3cm}
\begin{tabular}{l|lll}
cls (AUROC) & w/ masking & w/o masking & one seg \\ \hline
CoE         & 84.48\%    & 77.44\%     & 93.76\%
\end{tabular}
\end{table}

\subsection{Threats to Validity}

%\yuan{\sout{no threats to validity, something definitely need to discuss, either as subsection in Discussion or a full section}}

\noindent \textbf{Threats to Internal Validity.} Since CoE uses a two-level architecture as CATS~\cite{glavas_two-level_2020}, the inputs have more contexts compared to Transformer model and SentBERT, where they only have one window as the input.

\noindent \textbf{Threats to External Validity.} Although we have evaluated the performance of CoE by over 66,000 test samples, those segments only come from 500 files. The majority of the JavaScript Application bundles are not evaluated. In order to prove the generalizability of CoE, we need to build a complete manifest of SBoM for every third-party library with all versions posted on NPM.

\section{Related Work} \label{sec:related_work}

\subsection{The State of Software Bill of Materials}
SBoM is a considered as one of the best practices for software supply chain security by multiple institutions and governments~\cite{zahan2023software}. 
The growing number of supply chain attacks~\cite{hiesgen2022race, supply2021report, zahan2022weak} has underscored the necessity of SBoM.
However, recent studies have shown that integrating SBoM to existing software supply chain poses specific challenges, such as insufficient tooling~\cite{bi2023way, kloeg2024charting, zahan2023software}, lack of motivations~\cite{kloeg2024charting, zahan2023software, xia2023empirical}, time consumption~\cite{zahan2023software}, keeping SBoM up-to-date~\cite{balliu2023challenges, stalnaker2024boms}, and etc.
Most SBoM generation tools rely on dependency manifest such as Maven for Java and NPM for JavaScript~\cite{zahan2023software, rabbi2024sbom, nocera2023software}.
Thus, the SBoM for unknown supply chain software still remains an open challenge.

%\comment{fixed, we add all of the following literature. Most (all) of them are technical reports\yuan{because i am not familiar with SBoM, i start with this section. There are 58 citations for Xia et al's ICSE23 paper on SBoM, I briefly checked them and think these may be relevant to us? 1)Stalnaker, Trevor, et al. "Boms away! inside the minds of stakeholders: A comprehensive study of bills of materials for software systems." Proceedings of the 46th IEEE/ACM International Conference on Software Engineering. 2024. (ICSE'24), 2)Balliu, Musard, et al. "Challenges of producing software bill of materials for java." IEEE Security \& Privacy (2023). 3) Bi, Tingting, et al. "On the way to sboms: Investigating design issues and solutions in practice." ACM Transactions on Software Engineering and Methodology (2023). 4) Nocera, Sabato, et al. "Software bill of materials adoption: a mining study from GitHub." 2023 IEEE International Conference on Software Maintenance and Evolution (ICSME). IEEE, 2023. 5) Rabbi, Md Fazle, et al. "Sbom generation tools under microscope: A focus on the npm ecosystem." Proceedings of the 39th ACM/SIGAPP Symposium on Applied Computing. 2024. 6) Kloeg, Berend, et al. "Charting the Path to SBOM Adoption: A Business Stakeholder-Centric Approach." Proceedings of the 19th ACM Asia Conference on Computer and Communications Security. 2024. I may have time to read them and rewrite this paragraph a bit. It is of high chance that the reviewers are also not very familiar with work on SBOM}

\subsection{Text Segmentation and Classification}
Text segmentation and classification, sharing the same solution domain with SBoM, have been extensively studied and developed. The text segmentation and classification methods are well-suited for SBoM.
Recurrent neural networks (RNNs)~\cite{koshorek2018text, barrow2020joint} and Transformer architectures~\cite{glavas_two-level_2020, liu2021topic, lo2021transformer, inan2022structured} have demonstrated robust performance with better modeling of text or code semantics compared to static methods.
Moreover, the integration of text coherence modeling techniques~\cite{glavas_two-level_2020, lo2021transformer} into these deep learning models further optimizes the text segmentation task.
Additionally, the incorporation of pre-training processes~\cite{lo2021transformer, guo_graphcodebert_2021} enhances the understanding of complex context semantics and structures.

\subsection{Code Clone Retrieval}
The code clone retrieval task, which is to search code snippets with equivalent functionalities, can further enhance the precision of SBoM.
Recent advancements have leveraged deep neural networks~\cite{li2017cclearner, wei2018positive, zhao2018deepsim, zhang2019novel, yu2019neural, fang2020functional} to capture intricate code semantics for more accurate clone identification.
These deep learning-based methods often utilize hybrid features, such as Abstract Syntax Trees (ASTs)~\cite{zhang2019novel, yu2019neural}, encoded into continuous vector spaces to facilitate clone detection.
However, these techniques present specific limitations that need addressing.
Firstly, most deep learning-based approaches are optimized for exact matching or semantic similarity~\cite{fang2020functional}, overlooking the potential benefits of approximate nearest neighbors (ANN) algorithms.
Secondly, these methods often struggle when applied to extremely long code sequences~\cite{yu2019neural, fang2020functional}.
% This limitation imposes a challenge for both training and inference phases, as long sequences can be computationally expensive and may require specialized segmentation strategies to be effectively processed.

% \subsection{Multi-Task Learning}

% Multi-task learning (MTL) solves multiple learning tasks simultaneously while exploiting commonalities and differences across tasks~\cite{crawshaw2020multi}. This approach has gained significant momentum due to its potential to improve learning efficiency and prediction accuracy for individual tasks. The literature on MTL spans various domains, including natural language processing, computer vision~\cite{liu2019end, dai2016instance}, and bioinformatics, showcasing its versatility and broad applicability.

\section{Conclusion}
The Chain-of-Experts (CoE) framework is the first solution for SBoM generation from JavaScript application bundles, addressing the challenges of code segmentation, code classification, and code clone retrieval by an end-to-end deep learning model.
This integrated approach improves the accuracy and efficiency of the SBoM tasks.
% By leveraging the CoE framework, we unlock the potential to extend this methodology to encompass an even broader array of tasks, thereby enhancing the model's generalizability and applicability across different domains.
% The success of the CoE model in providing a comprehensive solution for JavaScript SBoM underscores the immense possibilities that lie in the strategic chaining of specialized tasks, paving the way for more sophisticated and versatile end-to-end architectures in the future.
We hope the idea of CoE's unified structure can provide an easier and comprehensive solutions for SBoM generation.

\bibliographystyle{IEEEtran}
\bibliography{cite}

\begin{thebibliography}{10}
\providecommand{\url}[1]{#1}
\csname url@samestyle\endcsname
\providecommand{\newblock}{\relax}
\providecommand{\bibinfo}[2]{#2}
\providecommand{\BIBentrySTDinterwordspacing}{\spaceskip=0pt\relax}
\providecommand{\BIBentryALTinterwordstretchfactor}{4}
\providecommand{\BIBentryALTinterwordspacing}{\spaceskip=\fontdimen2\font plus
\BIBentryALTinterwordstretchfactor\fontdimen3\font minus
  \fontdimen4\font\relax}
\providecommand{\BIBforeignlanguage}[2]{{%
\expandafter\ifx\csname l@#1\endcsname\relax
\typeout{** WARNING: IEEEtran.bst: No hyphenation pattern has been}%
\typeout{** loaded for the language `#1'. Using the pattern for}%
\typeout{** the default language instead.}%
\else
\language=\csname l@#1\endcsname
\fi
#2}}
\providecommand{\BIBdecl}{\relax}
\BIBdecl

\bibitem{xia2023empirical}
B.~Xia, T.~Bi, Z.~Xing, Q.~Lu, and L.~Zhu, ``An empirical study on software
  bill of materials: Where we stand and the road ahead,'' \emph{arXiv preprint
  arXiv:2301.05362}, 2023.

\bibitem{harutyunyan2020managing}
N.~Harutyunyan, ``Managing your open source supply chain-why and how?''
  \emph{Computer}, vol.~53, no.~6, pp. 77--81, 2020.

\bibitem{bi2023way}
T.~Bi, B.~Xia, Z.~Xing, Q.~Lu, and L.~Zhu, ``On the way to sboms: Investigating
  design issues and solutions in practice,'' \emph{ACM Transactions on Software
  Engineering and Methodology}, 2023.

\bibitem{stalnaker2024boms}
T.~Stalnaker, N.~Wintersgill, O.~Chaparro, M.~Di~Penta, D.~M. German, and
  D.~Poshyvanyk, ``Boms away! inside the minds of stakeholders: A comprehensive
  study of bills of materials for software systems,'' in \emph{Proceedings of
  the 46th IEEE/ACM International Conference on Software Engineering}, 2024,
  pp. 1--13.

\bibitem{kloeg2024charting}
B.~Kloeg, A.~Y. Ding, S.~Pellegrom, and Y.~Zhauniarovich, ``Charting the path
  to sbom adoption: A business stakeholder-centric approach,'' in
  \emph{Proceedings of the 19th ACM Asia Conference on Computer and
  Communications Security}, 2024, pp. 1770--1783.

\bibitem{zahan2023software}
N.~Zahan, E.~Lin, M.~Tamanna, W.~Enck, and L.~Williams, ``Software bills of
  materials are required. are we there yet?'' \emph{IEEE Security \& Privacy},
  vol.~21, no.~2, pp. 82--88, 2023.

\bibitem{supply2021report}
Sonatype, ``2021 state of the software supply chain report.''
  \url{https://www.sonatype.com/resources/state-of-the-software-supply-chain-2021},
  2021.

\bibitem{zahan2022weak}
N.~Zahan, T.~Zimmermann, P.~Godefroid, B.~Murphy, C.~Maddila, and L.~Williams,
  ``What are weak links in the npm supply chain?'' in \emph{Proceedings of the
  44th International Conference on Software Engineering: Software Engineering
  in Practice}, 2022, pp. 331--340.

\bibitem{rabbi2024sbom}
M.~F. Rabbi, A.~I. Champa, C.~Nachuma, and M.~F. Zibran, ``Sbom generation
  tools under microscope: A focus on the npm ecosystem,'' in \emph{Proceedings
  of the 39th ACM/SIGAPP Symposium on Applied Computing}, 2024, pp. 1233--1241.

\bibitem{kwaners0}
C.~Kwan, L.~Lindstr{\"o}m, D.~Giovannelli, K.~Podi{\c{n}}{\v{s}}, and
  D.~{\v{S}}trucl, ``Ers0: Enhancing military cybersecurity with ai-driven sbom
  for firmware vulnerability detection and asset management.''

\bibitem{cyclonedx}
CycloneDX, ``Owasp cyclonedx software bill of materials (sbom) standard,''
  \url{https://cyclonedx.org/}.

\bibitem{zhang2019novel}
J.~Zhang, X.~Wang, H.~Zhang, H.~Sun, K.~Wang, and X.~Liu, ``A novel neural
  source code representation based on abstract syntax tree,'' in \emph{2019
  IEEE/ACM 41st International Conference on Software Engineering (ICSE)}.\hskip
  1em plus 0.5em minus 0.4em\relax IEEE, 2019, pp. 783--794.

\bibitem{guo_graphcodebert_2021}
\BIBentryALTinterwordspacing
D.~Guo, S.~Ren, S.~Lu, Z.~Feng, D.~Tang, S.~Liu, L.~Zhou, N.~Duan,
  A.~Svyatkovskiy, S.~Fu, M.~Tufano, S.~K. Deng, C.~Clement, D.~Drain,
  N.~Sundaresan, J.~Yin, D.~Jiang, and M.~Zhou, ``{GraphCodeBERT}:
  {Pre}-training {Code} {Representations} with {Data} {Flow},''
  \emph{arXiv:2009.08366 [cs]}, Sep. 2021, arXiv: 2009.08366. [Online].
  Available: \url{http://arxiv.org/abs/2009.08366}
\BIBentrySTDinterwordspacing

\bibitem{8594636}
Y.~A. Malkov and D.~A. Yashunin, ``Efficient and robust approximate nearest
  neighbor search using hierarchical navigable small world graphs,'' \emph{IEEE
  Transactions on Pattern Analysis and Machine Intelligence}, vol.~42, no.~4,
  pp. 824--836, 2020.

\bibitem{reimers2019sentence}
N.~Reimers and I.~Gurevych, ``Sentence-bert: Sentence embeddings using siamese
  bert-networks,'' \emph{arXiv preprint arXiv:1908.10084}, 2019.

\bibitem{devlin2018bert}
J.~Devlin, M.-W. Chang, K.~Lee, and K.~Toutanova, ``Bert: Pre-training of deep
  bidirectional transformers for language understanding,'' \emph{arXiv preprint
  arXiv:1810.04805}, 2018.

\bibitem{vaswani_attention_2017}
\BIBentryALTinterwordspacing
A.~Vaswani, N.~Shazeer, N.~Parmar, J.~Uszkoreit, L.~Jones, A.~N. Gomez,
  L.~Kaiser, and I.~Polosukhin, ``Attention is {All} you {Need},'' in
  \emph{Advances in {Neural} {Information} {Processing} {Systems}},
  vol.~30.\hskip 1em plus 0.5em minus 0.4em\relax Curran Associates, Inc.,
  2017. [Online]. Available:
  \url{https://proceedings.neurips.cc/paper/2017/hash/3f5ee243547dee91fbd053c1c4a845aa-Abstract.html}
\BIBentrySTDinterwordspacing

\bibitem{shibata1999byte}
Y.~Shibata, T.~Kida, S.~Fukamachi, M.~Takeda, A.~Shinohara, T.~Shinohara, and
  S.~Arikawa, ``Byte pair encoding: A text compression scheme that accelerates
  pattern matching,'' 1999.

\bibitem{glavas_two-level_2020}
\BIBentryALTinterwordspacing
G.~Glavaš and S.~Somasundaran, ``\BIBforeignlanguage{en}{Two-{Level}
  {Transformer} and {Auxiliary} {Coherence} {Modeling} for {Improved} {Text}
  {Segmentation}},'' \emph{\BIBforeignlanguage{en}{Proceedings of the AAAI
  Conference on Artificial Intelligence}}, vol.~34, no.~05, pp. 7797--7804,
  Apr. 2020, number: 05. [Online]. Available:
  \url{https://ojs.aaai.org/index.php/AAAI/article/view/6284}
\BIBentrySTDinterwordspacing

\bibitem{beltagy2020longformer}
I.~Beltagy, M.~E. Peters, and A.~Cohan, ``Longformer: The long-document
  transformer,'' \emph{arXiv preprint arXiv:2004.05150}, 2020.

\bibitem{jiang2023low}
Z.~Jiang, M.~Yang, M.~Tsirlin, R.~Tang, Y.~Dai, and J.~Lin,
  ``“low-resource” text classification: A parameter-free classification
  method with compressors,'' in \emph{Findings of the Association for
  Computational Linguistics: ACL 2023}, 2023, pp. 6810--6828.

\bibitem{chen2018gradnorm}
Z.~Chen, V.~Badrinarayanan, C.-Y. Lee, and A.~Rabinovich, ``Gradnorm: Gradient
  normalization for adaptive loss balancing in deep multitask networks,'' in
  \emph{International conference on machine learning}.\hskip 1em plus 0.5em
  minus 0.4em\relax PMLR, 2018, pp. 794--803.

\bibitem{crawshaw2021slaw}
M.~Crawshaw and J.~Ko{\v{s}}eck{\'a}, ``Slaw: scaled loss approximate weighting
  for efficient multi-task learning,'' \emph{arXiv preprint arXiv:2109.08218},
  2021.

\bibitem{hiesgen2022race}
R.~Hiesgen, M.~Nawrocki, T.~C. Schmidt, and M.~W{\"a}hlisch, ``The race to the
  vulnerable: Measuring the log4j shell incident,'' \emph{arXiv preprint
  arXiv:2205.02544}, 2022.

\bibitem{balliu2023challenges}
M.~Balliu, B.~Baudry, S.~Bobadilla, M.~Ekstedt, M.~Monperrus, J.~Ron,
  A.~Sharma, G.~Skoglund, C.~Soto-Valero, and M.~Wittlinger, ``Challenges of
  producing software bill of materials for java,'' \emph{IEEE Security \&
  Privacy}, 2023.

\bibitem{nocera2023software}
S.~Nocera, S.~Romano, M.~Di~Penta, R.~Francese, and G.~Scanniello, ``Software
  bill of materials adoption: a mining study from github,'' in \emph{2023 IEEE
  International Conference on Software Maintenance and Evolution
  (ICSME)}.\hskip 1em plus 0.5em minus 0.4em\relax IEEE, 2023, pp. 39--49.

\bibitem{koshorek2018text}
O.~Koshorek, A.~Cohen, N.~Mor, M.~Rotman, and J.~Berant, ``Text segmentation as
  a supervised learning task,'' \emph{arXiv preprint arXiv:1803.09337}, 2018.

\bibitem{barrow2020joint}
J.~Barrow, R.~Jain, V.~Morariu, V.~Manjunatha, D.~W. Oard, and P.~Resnik, ``A
  joint model for document segmentation and segment labeling,'' in
  \emph{Proceedings of the 58th Annual Meeting of the Association for
  Computational Linguistics}, 2020, pp. 313--322.

\bibitem{liu2021topic}
J.~Liu, Y.~Zou, H.~Zhang, H.~Chen, Z.~Ding, C.~Yuan, and X.~Wang, ``Topic-aware
  contrastive learning for abstractive dialogue summarization,'' \emph{arXiv
  preprint arXiv:2109.04994}, 2021.

\bibitem{lo2021transformer}
K.~Lo, Y.~Jin, W.~Tan, M.~Liu, L.~Du, and W.~Buntine, ``Transformer over
  pre-trained transformer for neural text segmentation with enhanced topic
  coherence,'' \emph{arXiv preprint arXiv:2110.07160}, 2021.

\bibitem{inan2022structured}
H.~Inan, R.~Rungta, and Y.~Mehdad, ``Structured summarization: Unified text
  segmentation and segment labeling as a generation task,'' \emph{arXiv
  preprint arXiv:2209.13759}, 2022.

\bibitem{li2017cclearner}
L.~Li, H.~Feng, W.~Zhuang, N.~Meng, and B.~Ryder, ``Cclearner: A deep
  learning-based clone detection approach,'' in \emph{2017 IEEE International
  Conference on Software Maintenance and Evolution (ICSME)}.\hskip 1em plus
  0.5em minus 0.4em\relax IEEE, 2017, pp. 249--260.

\bibitem{wei2018positive}
H.~Wei and M.~Li, ``Positive and unlabeled learning for detecting software
  functional clones with adversarial training.'' in \emph{IJCAI}, 2018, pp.
  2840--2846.

\bibitem{zhao2018deepsim}
G.~Zhao and J.~Huang, ``Deepsim: deep learning code functional similarity,'' in
  \emph{Proceedings of the 2018 26th ACM Joint Meeting on European Software
  Engineering Conference and Symposium on the Foundations of Software
  Engineering}, 2018, pp. 141--151.

\bibitem{yu2019neural}
H.~Yu, W.~Lam, L.~Chen, G.~Li, T.~Xie, and Q.~Wang, ``Neural detection of
  semantic code clones via tree-based convolution,'' in \emph{2019 IEEE/ACM
  27th International Conference on Program Comprehension (ICPC)}.\hskip 1em
  plus 0.5em minus 0.4em\relax IEEE, 2019, pp. 70--80.

\bibitem{fang2020functional}
C.~Fang, Z.~Liu, Y.~Shi, J.~Huang, and Q.~Shi, ``Functional code clone
  detection with syntax and semantics fusion learning,'' in \emph{Proceedings
  of the 29th ACM SIGSOFT international symposium on software testing and
  analysis}, 2020, pp. 516--527.

\end{thebibliography}

\end{document}